# Individual Vanadium Dopants Form Deep In-Gap States in Monolayer $WS_2$

Tianhui Zhu[1 *], Carlos A. Gonzalez[1 *], Shihao Tu[3], Søren Tornøe[2,4], Ivan Pelayo[1], Dong-Rong Wu[5], Zhehao Ge[1,†], Hem Prasad Bhusal[1], Kenji Watanabe[7], Takashi Taniguchi[8], Nobuhiko P. Kobayashi [2,4], Yuan Ping[3,5,6, ‡], Jairo Velasco Jr.[1,2, ‡], Aiming Yan[1,2, ‡]

1. Department of Physics, University of California, Santa Cruz, CA, 95064, USA

2. Materials Science and Engineering Program, University of California, Santa Cruz, CA, 95064, USA

3. Department of Materials Science and Engineering, University of Wisconsin – Madison, Madison, WI, 53706, USA

4. Department of Electrical and Computer Engineering, Baskin School of Engineering, University of California Santa Cruz, Santa Cruz, CA, 95064, USA

5. Department of Chemistry, University of Wisconsin – Madison, Madison, WI, 53706, USA

6. Department of Physics, University of Wisconsin – Madison, Madison, WI, 53706, USA

7. Research Center for Electronic and Optical Materials, National Institute for Materials Science, 1-1 Namiki, Tsukuba 305-0044, Japan

8. Research Center for Materials Nanoarchitectonics, National Institute for Materials Science, 1-1 Namiki, Tsukuba 305-0044, Japan

[*] Tianhui Zhu and Carlos A. Gonzalez contributed equally to this work

[†]Present address: Department of Physics, University of California, Berkeley, Berkeley, CA, 94720, USA

[‡]Correspondence should be addressed to yping3@wisc.edu, jvelasc5@ucsc.edu, and aiyan@ucsc.edu

## Abstract

Point defects in atomically thin materials have a strong impact on physical properties and those that induce in-gap states are advantageous for quantum information science and engineering (QISE). However, dopant engineering consisting of well-controlled synthesis and robust identification of in-gap states is challenging. In this work, we addressed this challenge by first using finely tuned chemical vapor deposition to incorporate vanadium dopants into a monolayer $WS_2$ (V-$WS_2$). Next, we utilized a suite of scanned probe microscopy techniques to identify and characterize individual dopants. The latter included conductive atomic force microscopy (cAFM), low temperature scanning tunneling microscopy and spectroscopy (STM/STS), and scanning transmission electron microscopy and unambiguously revealed that vanadium dopants form deep in-gap states 0.35 eV above the valence band maximum in V-$WS_2$. Our experimental results are well supported by first principles calculations and taken together demonstrate that V-$WS_2$ is a promising platform for QISE applications.

Semiconducting two-dimensional (2D) transition metal dichalcogenides (TMDs) encompass a vast family of materials with significant promise for the next-generation electronics. They possess unique electronic, optical, and magnetic properties that can be leveraged for the next-generation transistors, sensors, and quantum devices [1,2]. Due to the large surface-to-volume ratio in 2D materials, point defects in these atomically thin films are expected to have a more significant impact on the material properties compared to their 3D bulk counterparts, and those with certain functionalities can be advantageous for quantum applications [3]. Among the different types of semiconducting TMDs, $WS_2$ stands out because of its comparatively large band gap [4], chemical stability under ambient conditions [5], and compatibility with large-scale synthesis methods such as chemical vapor deposition (CVD) [6–9]. W-based TMDs also possess larger spin-orbit coupling than Mo-based TMDs [10]. The combination of these advantages with careful dopant engineering can yield deep in-gap states that are promising for quantum information science and engineering (QISE) [11]. In addition to the aforementioned strengths $WS_2$ is also predicted to host large spin coherence time [12,13]. However, the realization of in-gap states via dopant engineering in $WS_2$ is absent.

In-gap states have been realized in $WS_2$ for defects made under specialized conditions such as ultrahigh vacuum, including sulfur vacancies [14] and cobalt substitutions for sulfur [15], but the methods employed in these previous works are unscalable. Previous theoretical and optical spectroscopy-based experiments that incorporated individual vanadium dopants in their synthesized $WS_2$ films reported defect resonances in the valence band as opposed to the desired in-gap states for QISE studies/applications [8,16,17]. So far, however, spatially resolved measurements of intentionally vanadium doped $WS_2$ (V-$WS_2$) films are lacking.

In this work, we use a comprehensive approach consisting of controlled material synthesis, several complementary atomically resolved characterization and spectroscopic probes, and first-principles calculations (Density-Functional Theory (DFT) with hybrid functionals and spin-orbit couplings (SOCs)) to identify and analyze the defects in intentionally doped monolayer $WS_2$ films. Importantly, we find unambiguously that deliberately incorporated V dopants form deep in-gap states in $WS_2$ monolayer sheets. This detection was enabled by using complementary scanning probe microscopies (SPMs) that provide direct chemical and spectroscopic identification with atomic resolution. We used CVD to synthesize large area $WS_2$ on $SiO_2$/Si substrates with flexible doping capabilities enabled by liquid phase precursors [8,9]. A representative CVD-grown single-layer $WS_2$ is shown in Fig. 1a.

Schematics of the spatially resolved characterization probes used in our study are presented in Figs. 1(b-d). These characterization methods include conductive atomic force microscopy (cAFM) (Fig. 1b), scanning tunneling microscopy and spectroscopy (STM/STS) (Fig. 1c), and scanning transmission electron microscopy (STEM) (Fig. 1 d). Overall, the combined use of these techniques establishes a fast and efficient feedback loop between synthesis and characterization, enabling synthesis efforts to be tuned quickly to obtain target sample dopant densities with a fast turnaround. For instance, because cAFM characterization is performed under ambient conditions, V dopant densities can be determined almost immediately after sample synthesis, in contrast to STM and STEM which require longer turnaround times. Notably, as we describe below, STEM and STM/STS were used to determine the V dopant signature in cAFM to establish our efficient characterization to synthesis feedback loop. Another advantage of incorporating cAFM in the feedback loop is that a high-quality cAFM sample can be directly used for STM/STS studies, significantly facilitating the STM/STS sample preparation process.

Two types of heterostructure devices were fabricated using dry transfer methods for SPM characterizations (see SI section S2). The first one consists of a monolayer $WS_2$ sheet on a monolayer graphene film that is supported by a thin hexagonal boron nitride (hBN) flake (Fig. 1b), while the second sample consists of a monolayer $WS_2$ supported by an exfoliated thin graphite (Fig. 1c). The graphite is critical in providing the conductive pathway to mitigate the Schottky barriers in 2D semiconductors that precludes low temperature electrical characterization [18,19], such as STM/STS. These samples undergo several post fabrication steps, such as high temperature annealing and AFM tip sweeping [20] to enhance sample surface cleanliness. See methods for details on these processes. An AFM topography image of a sample after post fabrication processing is shown in Fig. 1e. A root mean square (RMS) roughness of ~40 pm was achieved in the 200nm × 200nm area shown in the inset, which is optimal for SPM characterization.

First, we benchmark our comprehensive approach by studying a pristine $WS_2$ film without any intentional doping. The use of cAFM provides quick characterization on defect density of a sample. Additionally, as we discuss below, the use of STEM and STM/STS aids in benchmarking the signatures and densities for intrinsic defects in $WS_2$, which will be utilized for analyzing growth and characterization of doped samples. Figure 2a presents a typical atom-resolved cAFM current map at a sample bias of $V_S = +200$ mV with two types of defects, one bright and one dark. Both defects have a diameter below 1 nm. At low biases of $|V_S| < 1$V, the accessible energy level is within the band gap of the $WS_2$ layer, and the local conduction can be understood by a direct tunneling model [21], where charge carriers directly tunnel between the AFM tip and the graphene layer in the substrate. These small bright and dark defects are seen throughout the sample. Their densities are close in value, around $(5 \pm 1) \times 10^{12}$ cm$^{-2}$ with region-to-region variations. A larger area cAFM scan at $V_S = +100$ mV in Fig. 2b shows the small defects with a density of ~ $(3.5 \pm 0.5) \times 10^{12}$ cm$^{-2}$, as well as a third type of defects with sizes on the order of 5 nm and a density of ~$1.6\times10^{11}$ cm$^{-2}$. The density for the former is probably an underestimation due to the limited resolution with clusters of defects, which explains the small discrepancy in the estimated density for Fig. 2a.

Based on the prevalence of the small defects over the entire 50nm × 50nm scan area in Fig. 2b, they are likely oxygen substitutions at the S sites, $O_S$ defects, which have been shown to be the dominant defect in as synthesized $WS_2$ samples [22]. The density of the larger defects is more than one order of magnitude lower than that of the oxygen defects. Previous STM studies of pristine TMDs have reported similar defect signatures with density much lower than $O_S$ defects and they were attributed to negatively charged hydrocarbon substitutions on the chalcogen sites, such as $CH_S^-$ [23–25]. Since these charged defects lower local conduction more than the $O_S$ defects and dominate the image contrast, the only other type of visible defect is the small bright $O_S$ defect in Fig. 2b. The exact nature of these defects cannot be determined based on cAFM data alone. In the discussion to follow, we use STM/STS in corroboration with DFT calculations and previous literature identify the different defect signatures resolved in cAFM.

STM can further elucidate the nature of as-grown $WS_2$ intrinsic defects by probing the local electronic properties. $O_S$ defects are the most frequently observed defects in the sample and are confirmed by comparing the STM topography image at $V_S = +1.1$ V in Fig. 2c with prior studies [14,22,23,26]. At this bias, the $O_S$ defects in both the top S layer ($O_S$ top) and the bottom S layer ($O_S$ bottom) have six-fold symmetry with a dark center surrounded by six dark spots [22,23]. The difference is that $O_S$ bottom has an extra bright ring surrounding the dark center [22,23]. We observe six $O_S$ top defects and one $O_S$ bottom defect in this 10nm × 10nm window, which corresponds to a total defect density of about $7\times10^{12}$ $cm^{-2}$. This value is comparable to the reported $O_S$ defect density of around $5\times10^{12}$ $cm^{-2}$ in CVD grown TMDs [23,26].

By comparing the contrasts of $O_S$ defects relative to the background in Fig. 2a to the STM results, we can assign the small dark defects as $O_S$ top and the small bright defects as $O_S$ bottom in cAFM measurements. We note that although both cAFM and STM scans present similar total density of $O_S$ defects, cAFM scan has equal numbers of $O_S$ top and $O_S$ bottom defects while STM scan shows significantly more $O_S$ top defects. This could be a direct result of the different fabrication processes as detailed in methods section. The STM sample was fabricated from top down, meaning that the exposed top surface is the same as the top $WS_2$ surface during CVD growth. The cAFM device was fabricated from bottom up using the flip chip method and the top surface was originally the bottom surface in contact with the $SiO_2$/Si chip during growth. A similar top-down approach observed an increase in top $O_S$ defects on the $WS_2$ sample [26], while a study on $WSe_2$ reveals that the flip chip transfer did not affect the density of $O_{se}$ defects [27]. Both studies agree with our observations. This suggests that certain post-growth processes contribute to the formation of extra oxygen defects on the exposed surfaces.

Representative $\mathrm{d}I/\mathrm{d}V_S$ spectra at the defect center and away from the defect taken in two separate scanning sessions are shown in Fig. 2(e-f) for $O_S$ top and $O_S$ bottom respectively. Following Ref. [28], we take the logarithm of the $\mathrm{d}I/\mathrm{d}V_S$ values and reveal that the valence band maximum (VBM) occurs at around $V_S = (-1550 \pm 50)$ mV and the conduction band minimum (CBM) is at around $V_S = (950 \pm 50)$ mV (see SI section S5). The band gap is estimated to be approximately $(2.5 \pm 0.1)$ eV, larger than that of $WSe_2$ [29,30]. Notably, both types of $O_S$ defects in our data show no in-gap states, only defect resonances ~ 200—300 meV below the VBM. This agrees with prior STM/STS results [22,23].

Our experimental results are further supported by hybrid-functional DFT calculations including SOC, with the fraction of Fock exchange fixed by Koopmans' condition [31] (See SI section S6). The calculated density of states for $WS_2$ monolayers in the pristine form and with an isolated $O_S$ defect are shown in Fig. 2f. Compared with pristine $WS_2$, the $O_S$ defect does not introduce any localized states inside the band gap, and the band gap size and band edge states remain unchanged. Instead, there is enhanced density of states (DOS) at about 0.6 eV below the VBM. This is attributed to a defect resonance arising from hybridization with the host valence-band states, in agreement with Ref. [32]. The DFT results qualitatively reproduce the key features observed in the STS measurements in Fig. 2(d-e), namely the preservation of the band gap and the emergence of resonances inside the valence band near the VBM. Furthermore, the defect formation energy as a function of Fermi level (see SI section S7) indicates that the thermodynamically stable charge state of an isolated $O_S$ defect is neutral when the Fermi level is within the band gap, and no charge transition levels (CTLs) are found within the band gap. This is consistent with the absence of in-gap states in both the calculated DOS and the experimental STS data, as well as the absence of band bending in STS for a neutral defect.

Identification of the intrinsic defects in pristine $WS_2$ establishes a baseline for our study of intentionally doped $WS_2$ samples. In the following, we present the characterization of V-$WS_2$, using optical, STEM and SPM measurements in corroboration with DFT calculations. We note that V concentrations in the CVD growth precursors do not directly translate to atomic concentrations in the final products and we use atomic concentrations from STEM data to more accurately describe the dopant concentrations in our samples.

Monolayer V-$WS_2$ flakes were successfully grown on $SiO_2$ with triangular domains larger than 50 μm on each side. V concentrations in the precursor solution affect the appearance of the flakes formed on the substrates. In the case of 50% V precursor, as shown in Fig. 3a, there is a higher number of flakes overall but many of them are either small in lateral area or not uniform in thickness. The larger flakes tend to connect to each other. In comparison, 1% V precursor yields more uniform and isolated monolayer flakes in Fig. 3b, which is preferred for SPM characterization.

Although not 1:1 translation in V concentration, higher concentration of V precursor yields higher atomic density of V dopants in the final products. This correlation can be captured by optical spectroscopy. A higher V concentration in the precursor results in a decrease of 2LA(M) mode and an increase of the defect-activated LA(M) and ZA(M) modes in the Raman data shown in Fig. 3c as well as a redshift of the photoluminescence (PL) peak and a quenching of the intensity in Fig. 3d. All above spectroscopic features indicate a higher atomic density of V dopants. However, once the dopant precursor concentration is below a threshold value of 5%, there are minimal changes in the optical measurements (see SI section S8). This is likely due to the spot size of the laser (approximately 1 μm), where the signal from the pristine regions saturates the LA(M) and ZA(M) vibrational modes activated from the low V dopant density. Atomically resolved techniques are thus necessary to more accurately measure the dopant density in samples with dilute dopant concentrations, which can be achieved by combining cAFM and STEM as shown below.

Continuing to the dilute V-$WS_2$ spatially resolved characterization, we expect to see oxygen impurities, the most prevalent defects observed in pristine $WS_2$, as well as the V dopants and $CH_S^-$ defects. Figure 3e shows a cAFM current map of the V-$WS_2$/graphite sample at $V_S$= +50 mV. The small bright and dark defects present in the pristine $WS_2$ samples are also observed in the V-$WS_2$. A new feature is also observed, seen as a dark spot similar to the smaller dark defect, but with a larger diameter of approximately 3-5 nm. The small bright defects have a density of around $4\times10^{12}$

cm$^{-2}$ and correspond to the $O_S$ bottom defects, because they match the density in the pristine case above. Since the large dark spot defects have greater contrast here, the $O_S$ top defects are harder to discern for an estimation of their density. We observe a density of about $(2.5 \pm 0.5) \times 10^{12}$ cm$^{-2}$ for the large dark spot defects. The uncertainty is due to difficulty in differentiating the defects when they cluster. SI sections S9 and S10 present additional cAFM images which reveal how these defects suppress conductance at both positive and negative bias.

STEM images of V-$WS_2$ lattices with similar growth conditions to the cAFM sample above are shown in Fig. 3(f-g). The intensity of high-angle annular dark-field (HAADF)-STEM image is roughly proportional to the square of the atomic number [33], which means the contrasts are largely dominated by W atoms. V atoms ($Z$ = 23) substitute for W atoms ($Z$ = 74) and form individual $V_W$ defects that are darker in contrast at lattice sites and are nanometers apart. This window shows a doping concentration as low as 0.15% in V atomic concentration or $2 \times 10^{12}$ cm$^{-2}$ in density, which is consistent with the large dark spot defect density measured by cAFM. Because of this agreement, we conclude that the identity of the large dark spot defects is the V dopants. A zoom-in of the outlined region in Fig. 3f is provided in Fig. 3g. The bright atoms are the W atoms in hexagonal lattices, and they appear darker when substituted by V atoms as expected. The S site defects are less observable due to the lower atomic numbers.

The STEM images allow us to precisely measure the density of V doping, providing further confirmation with the cAFM results and indicating that most of the defects seen in cAFM are not the $CH_S^-$ defects. This is especially helpful with dilute dopants as optical characterization is incapable of distinguishing samples with precursor concentrations below 5%. Even though $V_W$ and $CH_S^-$ defects lack distinctive features with cAFM characterization, STEM experiments (and STM experiments described below) show that the number of the occasional $CH_S^-$ defects should not exceed 10% of that of the V dopants, which is within the estimated uncertainty of the $V_W$ density measured by cAFM. By scanning and cross-checking densities at different regions, we minimize impact of small variations on the spatial distribution of the defects. Additional discussion on the distinction between the $V_W$ and the $CH_S^-$ defects can be found in the STM experiment section to follow and SI section S11.

Now that the presence of isolated V dopants in our $WS_2$ films is well established and a dilute concentration limit was achieved, we use STM/STS to directly reveal electronic signatures of $V_W$ defects in $WS_2$. We observe several large dark defects in the STM topography image at $V_S = +2V$, as shown in Fig. 4a. Their density is $\sim 3\times10^{12}$ $cm^{-2}$ in this window, consistent with the STEM and cAFM characterizations for vanadium. Most of other scanned regions also show dominance of these large dark defects with similar densities. We thus robustly conclude that these large defects are $V_W$ defects. Figure 4(b-c) present the STM topography images of one such defect at $V_S = -2.5V$ and $V_S = +2$ V. The defect contrast changes from bright to dark as the bias changes from within the valence band to within the conduction band. This trend is similar to our cAFM observations. Other than the large $V_W$ defects, there are smaller bright and dark defects, which present the same characteristics as the $O_S$ defects previously observed in the pristine sample and are consistent with cAFM observation on V-$WS_2$.

Next, we focus on STS of an isolated V dopant. Comparing the $dI/dV_S$ spectra taken on the defect circled in Fig. 4a (Fig. 4d) and at 3 nm away from the defect (Fig. 4e) shows unambiguously that this defect has an induced in-gap state close to the VBM. There is a shift of the conduction band edge towards more positive biases on the V dopant, which was not observed for the neutral $O_S$ defects in Fig. 2(d-e). This indicates that the defect is negatively charged and this upward band bending behavior has been reported previously for other negatively charged defects [23,25,34,35].

Inset in Fig. 4d was taken over a smaller bias window with an AC excitation of 1mV to reveal that the in-gap states consist of one sharp peak close to $V_S = -1200$ mV followed by a few less pronounced peaks. The deep in-gap state of the V dopant is mapped in Fig. 4f, which shows a three-fold symmetry. Notably, both the $dI/dV_S$ spectra and the $dI/dV_S$ map of the in-gap states differentiate $V_W$ from $CH_S^-$ – the in-gap states of $CH_S^-$ show up as two small peaks followed by one large peak in STS and the mapping of its deepest in-gap state has more complicated structures. The details of these features are shown in SI section S11. Importantly, our STM measurements reveal that the density of the $CH_S^-$ defects is one order of magnitude lower than that of the $V_W$ defects and thus has minimal contribution to the V dopant count.

Computing $V_W$ defect formation energy and thermodynamic CTL gives a (0/−1) charge transition level at 0.77 eV above the VBM, which indicates a deep p-type defect as shown in Fig. 4g. When the Fermi level lies between this CTL and the CBM, the defect is expected to be negatively charged. This agrees with the upward band bending observed in Fig. 4(b-c). As discussed in Ref. [28,33], the quasiparticle energy of a defect state (i.e., its position in the single-particle band structure) is physically equivalent to the thermodynamic charge transition level (CTL), up to the lattice relaxation energy associated with the change in charge state. We find that the geometry relaxation energy of charged state is negligible (only 0.02eV), and therefore the gap state is at 0.75 eV above VBM. This is also consistent with the isolated gap state in the DOS calculation. We note that DFT calculations estimate the defect level to be deeper in the gap (0.75 eV above the VBM) compared to STS measurements (~0.35 eV above VBM), which could be a result of a substrate effect in experiments. As discussed in Ref. [36], the quasiparticle gap and ionization energies of defect states are reduced upon the addition of a substrate or with increasing thickness of the 2D material, owing to enhanced dielectric screening. These in-gap states are absent in pristine $WS_2$ and originate predominantly from the V 3d orbitals. The corresponding defect state is predominantly composed of V $d_z^2$ orbitals and thus belongs to the a1' representation of the local $D_{3h}$ point group. The real-space wavefunction in Fig. 4h further illustrates the strong localization and the threefold symmetry of this defect state, which resembles a threefold symmetry evident in the experimental $dI/dV_S$ maps in Fig. 4f. This provides a direct microscopic origin for the experimentally observed in-gap states associated with $V_W$.

By use of several atomically resolved imaging and spectroscopic techniques, we have clearly identified intentional V dopants in monolayer $WS_2$ and discovered that they introduce occupied deep in-gap states. These states are desirable for future QISE applications [11] but were not observed in previous studies of V-$WS_2$ that mainly utilized optical spectroscopy [8,16,17]. Our discovery was enabled by low temperature STM/STS, which provided atomic scale identification followed by direct spectroscopic evidence of in-gap states at about 0.35 eV above VBM for isolated V dopants. In addition to STM/STS, we used cAFM and STEM to facilitate defect identification and guidance during material synthesis optimization. Moreover, the experimental observations pertaining to identification of prominent defect types, such as $O_S$ and $V_W$, were further bolstered by high level first-principles calculations at hybrid-functional DFT with spin-orbit couplings. Besides the discovery of a new deep in-gap state in $WS_2$, we also demonstrated cAFM can serve as a quick and efficient tool to provide feedback on material synthesis and dilute dopant incorporation. This will help achieve faster turnaround time for future CVD synthesis of doped $WS_2$ and other TMD samples for QISE applications.

## Acknowledgements

We thank S. Borggreve and H. Zandvliet for their helpful discussions on cAFM practices and assistance at the initial stage of developing cAFM technique for this project. We also thank D. Lederman for assistance with the cryogenic recovery system. A.Y. and C.A.G acknowledge support from W. M. Keck Foundation. A.Y. acknowledges support from NSF award 2235474. J.V.J., T.Z., H.P.B., C.A.G, D.W., and Y.P. acknowledge support from the Gordon and Betty Moore foundation award 10.37807/GBMF11569. J.V.J. acknowledges support from NSF award 2403491 and from UCSC Office of Research SEED award. K.W. and T.T. acknowledge support from the CREST (JPMJCR24A5), JST and World Premier International Research Center Initiative (WPI), MEXT, Japan. STEM was performed at nano@stanford RRID:SCR_026695. C.A.G and A.Y. acknowledge technical support from Dr. Pinaki Mukherjee at Stanford Nano Shared Facilities. S. Tu. and Y.P. also acknowledge the support from NSF award under grant no. DMR-2143233. This portion of the work used the TACC Stampede3 system at the University of Texas at Austin through allocation PHY240212 from the Advanced Cyberinfrastructure Coordination Ecosystem: Services and Support (ACCESS) program [37], which is supported by US National Science Foundation grants No. 2138259, No. 2138286, No. 2138307, No. 2137603, and No. 2138296.

## Author contributions

A.Y. and J.V.J., conceived the project. A.Y., J.V.J., T.Z. and C.A.G. designed the research strategy. C.A.G. synthesized the TMD samples under the supervision of A.Y. T.Z. and C.A.G. fabricated the devices for scanning probe measurements with assistance from I.P. and under the supervision of J.V.J. and A.Y. T.Z. and C.A.G. performed cAFM measurements and analysis under the supervision of J.V.J. and A.Y. T.Z. performed STM measurements and analysis with assistance from I.P. and Z.G. and under the supervision of J.V.J. S.Tu performed DFT calculations with assistance from D.R.W. and under the supervision of Y.P. C.A.G. performed STEM measurements and analysis with assistance from H.P.B. and under the supervision of A.Y. S.Tornø carried out optical measurements under the supervision of N.K. K.W. and T.T. provided hBN crystals. T.Z., C.A.G., J.V.J., and A.Y. wrote the manuscript with feedback from all authors.

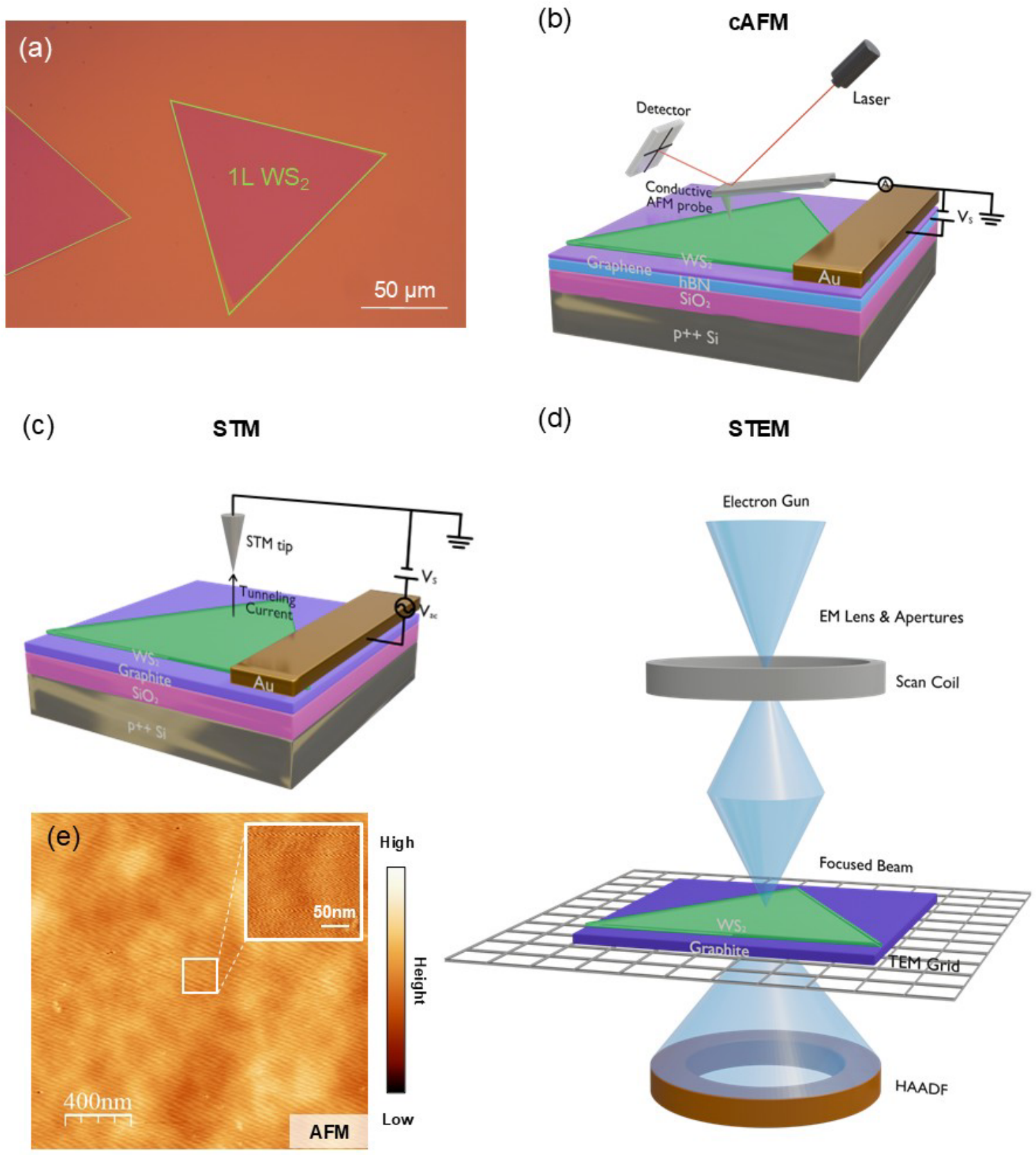

**Figure 1 Overview of characterization techniques and transition metal dichalcogenide (TMD) devices. (a)** Optical image of pristine $WS_2$ grown on $SiO_2$/Si substrate with flake edge lengths longer than 50 µm. Schematics of atomically resolved imaging techniques including **(b)** conductive atomic force microscopy (cAFM), **(c)** scanning tunneling microscopy (STM) and **(d)** scanning transmission electron microscopy (STEM). Two different types of heterostructures are shown: one is $WS_2$ on graphene and hexagonal boron nitride (hBN) (b), and the other is $WS_2$ supported on graphite (c and d). **(e)** AFM topography scans on a clean $WS_2$/graphene/hBN heterostructure. Inset shows a 200 nm × 200 nm region with root mean square roughness of ~ 40pm.

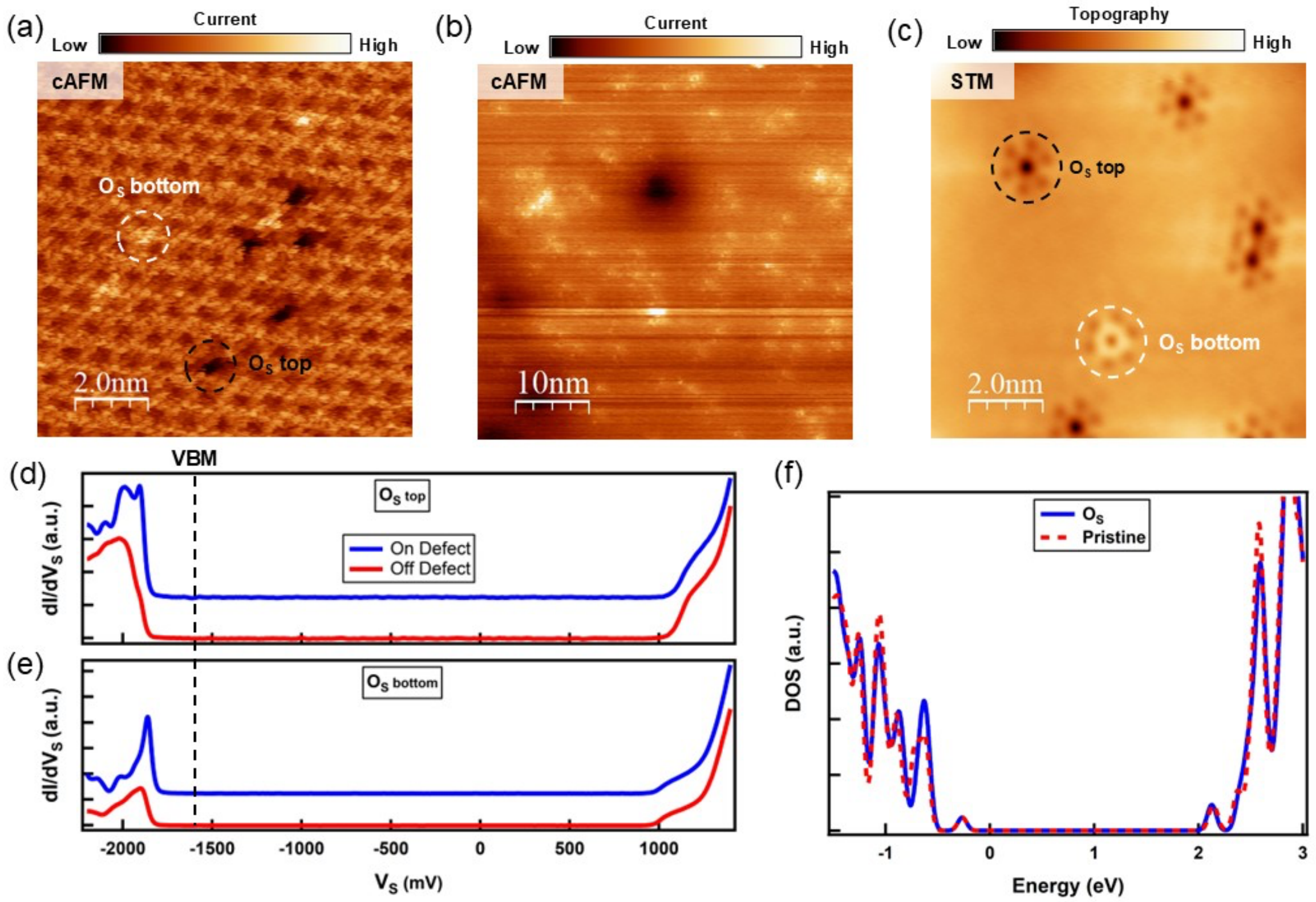

**Figure 2 Characterization of pristine $WS_2$ monolayer with atomic resolution, using scanning probe microscopies. (a)** Atomically resolved $WS_2$ lattices via cAFM at $V_S = +200$ mV showing two types of defects with bright and dark contrasts. **(b)** Large area scan with cAFM at $V_S = +100$ mV presents a third type of defect, which is much larger in size with dark contrast. **(c)** STM topography scan showing both $O_S$ top (dark center) and $O_S$ bottom (dark center with bright ring) defects. **(d-e)** $dI/dV_S$ spectra of $O_S$ top and $O_S$ bottom defects, respectively. Both defect spectra are compared to spectra on pristine $WS_2$ away from defects. (f) DFT calculated density of states (DOS) versus energy on (blue) and off (red) a single $O_S$ defect, showing no in-gap states. The Fermi level is set to zero energy.

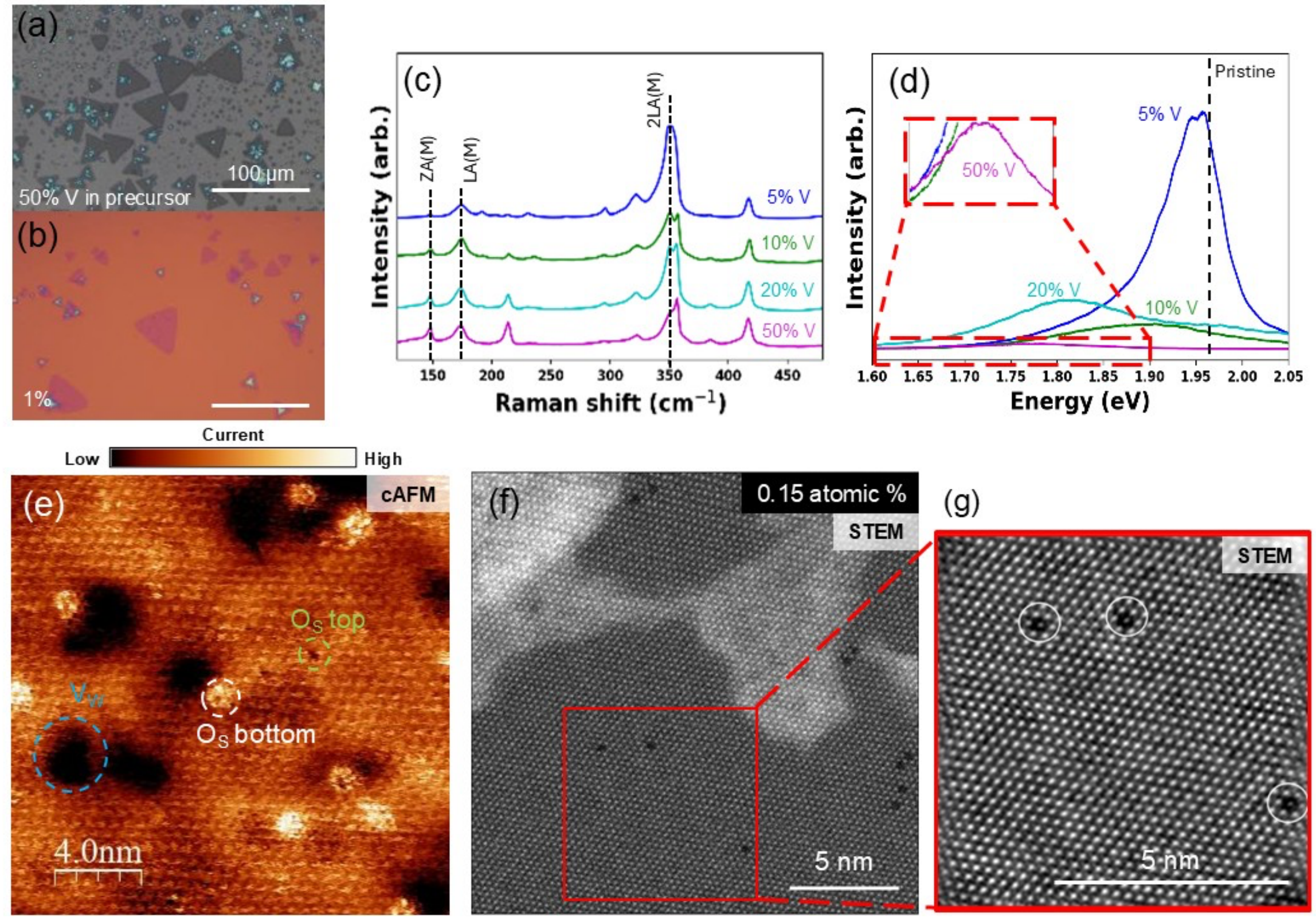

**Figure 3 Characterization of V-$WS_2$ with SPM and optical spectroscopy. (a-b)** Optical images of growths with (a) 50% and (b) 1% V concentrations in precursor solutions show more flakes with higher V concentrations. **(c-d)** Raman and PL of concentrations from 5% to 50%, respectively. We note that precursor concentrations do not directly translate to atomic concentrations. **(e)** cAFM current image at $V_S$ = +50 mV shows $V_W$ defects (dashed blue circle) as large dark defects, in addition to the $O_S$ defects (dash white and green circles) as seen previously in the pristine sample. **(f)** HAADF-STEM image identifies the presence of atoms lighter than the W atoms at the W sites, with atomic concentration of around 0.15%. **(g)** FFT-filtered zoomed-in image of outlined region in (f), clearly showing three W site defects.

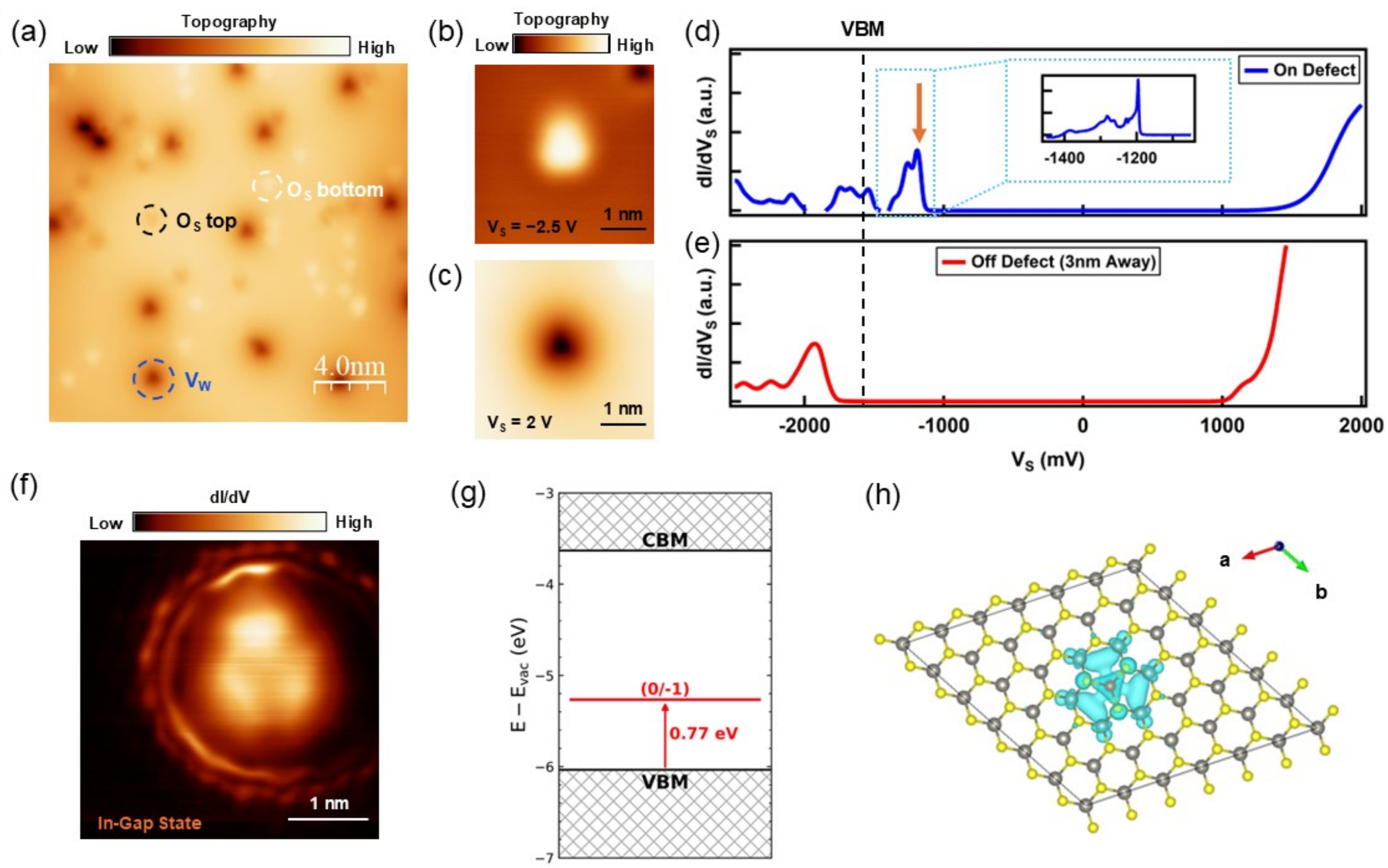

**Figure 4 STM/STS characterization and DFT modeling of V-$WS_2$. (a)** STM topography at $V_S$ = 2V, where large dark depressions correspond to V dopants. The V dopant exhibits different features in topographic maps at varying samples biases, going from **(b)** bright to **(c)** dark for $V_S = -2.5$V and $V_S$ = 2V respectively. d$I$/d$V_S$ spectra **(d)** of the $V_W$ defect circled in (a) and (c) at a defect-free spot 3nm away **(e)**, showing defect-induced in-gap states near the VBM and upward shifting of the CBM as a result of negative charge. Inset in (d) zooms in on the in-gap states to show one sharp peak followed by several smaller peaks. AC excitation is 20mV for (d) and (e) and is 1mV for inset of (d). **(f)** A representative d$I$/d$V_S$ map of the $V_W$ defect at the deepest in-gap state energy, exhibiting three-fold symmetry. **(g)** Thermodynamic charge transition level of defect $V_W$ (0/-1) with SOC included. **(h)** Calculated real-space distribution of the in-gap state wave function amplitude.

# Supplementary Information

## Individual Vanadium Dopants Form Deep In-Gap States in Monolayer $WS_2$

Tianhui Zhu[1 *], Carlos A. Gonzalez[1 *], Shihao Tu[3], Søren Tornø[2,4], Ivan Pelayo[1], Dong-Rong Wu[5], Zhehao Ge[1,†], Hem Prasad Bhusal[1], Kenji Watanabe[7], Takashi Taniguchi[8], Nobuhiko P. Kobayashi [2,4], Yuan Ping[3,5,6, ‡], Jairo Velasco Jr.[1,2, ‡], Aiming Yan[1,2, ‡]

1. Department of Physics, University of California, Santa Cruz, CA, 95064, USA

2. Materials Science and Engineering Program, University of California, Santa Cruz, CA, 95064, USA

3. Department of Materials Science and Engineering, University of Wisconsin – Madison, Madison, WI, 53706, USA

4. Department of Electrical and Computer Engineering, Baskin School of Engineering, University of California Santa Cruz, Santa Cruz, CA, 95064, USA

5. Department of Chemistry, University of Wisconsin – Madison, Madison, WI, 53706, USA

6. Department of Physics, University of Wisconsin – Madison, Madison, WI, 53706, USA

7. Research Center for Electronic and Optical Materials, National Institute for Materials Science, 1-1 Namiki, Tsukuba 305-0044, Japan

8. Research Center for Materials Nanoarchitectonics, National Institute for Materials Science, 1-1 Namiki, Tsukuba 305-0044, Japan

[*] Tianhui Zhu and Carlos A. Gonzalez contributed equally to this work

[†]Present address: Department of Physics, University of California, Berkeley, Berkeley, CA, 94720, USA

[‡]Correspondence should be addressed to yping3@wisc.edu, jvelasc5@ucsc.edu, and aiyan@ucsc.edu

## S1. Synthesis of $WS_2$ monolayers

Monolayer vanadium-doped tungsten disulfide (V-$WS_2$) flakes were synthesized via chemical vapor deposition (CVD) with a liquid transition metal solution precursor. Sodium tungstate ($Na_2WO_4$) and sodium metavanadate ($Na_2VO_3$) were diluted in deionized water, 25 mM and 1.5-25 mM respectively. They were then mixed with varying ratio of V:W precursors to control the doping level of V in $WS_2$ flakes. The solution was spin-coated on 285 nm $SiO_2$/Si substrates and loaded into a tube furnace with S powder (200 mg) upstream. The CVD growth was performed by heating the S powder to 220 °C and the spin-coated substrates to 850 °C for 15 min with a gas flow mixture (Ar = 90 sccm, $H_2$ = 10 sccm). The furnace was then allowed to cool naturally after the CVD growth.

To grow pristine $WS_2$ flakes, a similar liquid solution process was followed without the V precursor. The CVD growth time was reduced to 10 minutes and there was no $H_2$ introduced in the carrier gas flow.

The pristine $WS_2$ sample for scanning tunneling microscopy (STM) study was prepared with a transition metal powder precursor mixture rather than liquid solution. $WO_3$ (30 mg) and NaCl (6.5 mg) were mixed to make the transition metal precursor. For the growth of $WS_2$ monolayer flakes, the S powder was heated to 250 °C and transitional metal precursor mixture to 830 °C, maintained for 7 min.

## S2. Device Fabrication

Thin graphite, graphene and hexagonal boron nitride (hBN) were mechanically exfoliated onto Si chips with 285nm thick $SiO_2$ using Scotch tape. The pristine $WS_2$ on graphite device for STM was fabricated using polyethylene terephthalate (PET) by picking up $WS_2$ flake from $SiO_2$ and dropping it down on an exfoliated graphite flake. The other samples were fabricated using polypropylene carbonate (PPC) and a modified version of the flip chip method [1] by picking up bottom flakes first and the transition metal dichalcogenide (TMD) flake last. Compared to exfoliated TMD flakes, CVD grown flakes adhere to growth substrate more strongly. A drop of deionized water was used to help delaminate flakes from the $SiO_2$ substrate as needed [2]. After the pick-up, PPC film with the stack on top was peeled off from the PDMS stamp and placed on a clean Si chip with the PPC side down. The exposed TMD top surface never directly contacted polymers, leading to a cleaner device surface.

In order to remove polymer residue, all samples were annealed in forming gas (10%-20% $H_2$/Ar) at 350 °C for scanning probe microscopy (SPM) samples and at 300 °C for scanning transmission electron microscopy (STEM) samples. Subsequently, for scanning probe samples, an atomic force microscopy (AFM) tip (BudgetSensors, ContGD-G) was used to continuously sweep the sample surfaces in contact mode for > 10 hrs and a surface roughness of less than 100 pm across a 200 nm × 200 nm window was achieved. Electrodes (5nm Cr/60nm Au) were deposited though a shadow mask using a thermal evaporator (vacuum pressure < $1\times10^{-6}$ mbar).

## S3. Experimental characterization methods

### Raman and photoluminescence (PL)

Raman and PL measurements were performed using 532 nm excitation laser in a HORIBA Scientific XploRA PLUS Raman spectroscope.

### STEM

High Angle Annular Dark Field (HAADF)-STEM was performed in a Thermo Fisher Spectra TEM at the Stanford Nano Shared Facilities, operating at 80kV. A Gaussian blur filter was applied to reduce noise and enhance the visibility of atomic features.

### SPM

STM measurements were performed in a CreaTec scanning tunneling microscope under ultrahigh vacuum (UHV, $\sim 5\times 10^{-11}$) at 5K. A chemically etched tungsten tip was calibrated on Au (111) prior to sample measurements. STM samples were annealed in UHV at 350 °C for about 8 hours before loading into the STM head.

AFM measurements were performed by an Asylum Research Cypher S inside a $N_2$-filled glovebox. For conductive AFM, Pt coated probes (OPUS by MikroMasch, 240AC-PP) and conductive diamond coated probes (BudgerSensors, AIO-DD) were used.

Biases were applied to the samples in STM and cAFM measurements while the tips were grounded. WSxM software [3] was used to analyze STM and AFM images.

## S4. Density functional theory

First-principles calculations were mainly performed using the open-source plane-wave DFT package Quantum ESPRESSO [4]. Fully relativistic norm-conserving pseudopotentials from the PseudoDojo library [5] were employed, with a plane-wave energy cutoff of 70 Ry. The pristine monolayer $WS_2$ structure was optimized using the Perdew-Burke-Ernzerhof (PBE) exchange–correlation functional [6], yielding a lattice constant of a=3.187 Å. The structure of the defect supercells was optimized until the residual forces were below $5\times10^{-4}$ Ry/Bohr, and the corresponding SCF calculations were converged to $10^{-7}$ Ry in total energy. A vacuum spacing of 35 Bohr was applied to avoid spurious interlayer interactions.

To achieve a more accurate description of the monolayer $WS_2$ band gap, hybrid-functional calculations were performed using the PBE0(α) functional with spin-orbit couplings. The fraction of exact exchange α is determined by enforcing the generalized Koopmans' condition [7], which requires the electron affinity of the (q+1) charge state to equal the ionization potential of the q charge state, i.e., $EA_{q+1}=IP_q$. In periodic supercell calculations, charged defects artificially interact with their periodic replicas and homogenous background counter-charge. Therefore, charged cell corrections $E_{CCC}$ were applied to the charged defect calculation of both total energies and eigenvalues (see Ref. [8,9] for a detailed description of the theoretical formalism and implementation).

Density-of-states calculations including noncollinear spin and spin–orbit coupling for the defective supercells were carried out using the plane-wave pseudopotential code Vienna ab initio simulation package (VASP) [10,11], with an energy cutoff of 500 eV.

## S5. Determination of band gap in $WS_2$

The band gap in $WS_2$ is determined via scanning tunneling spectroscopy (STS) by taking the logarithm of the dI/dV spectra following Ref. [12]. As shown in Fig. S1, the valence band maximum (VBM) occurs at around $V_S = (-1550 \pm 50)$ mV and the conduction band minimum (CBM) is at around $V_S = (950 \pm 50)$ mV. The band gap is about $(2.5 \pm 0.1)$ eV. This observation is consistent at locations away from defects on both pristine and vanadium-doped $WS_2$ (V-$WS_2$) samples.

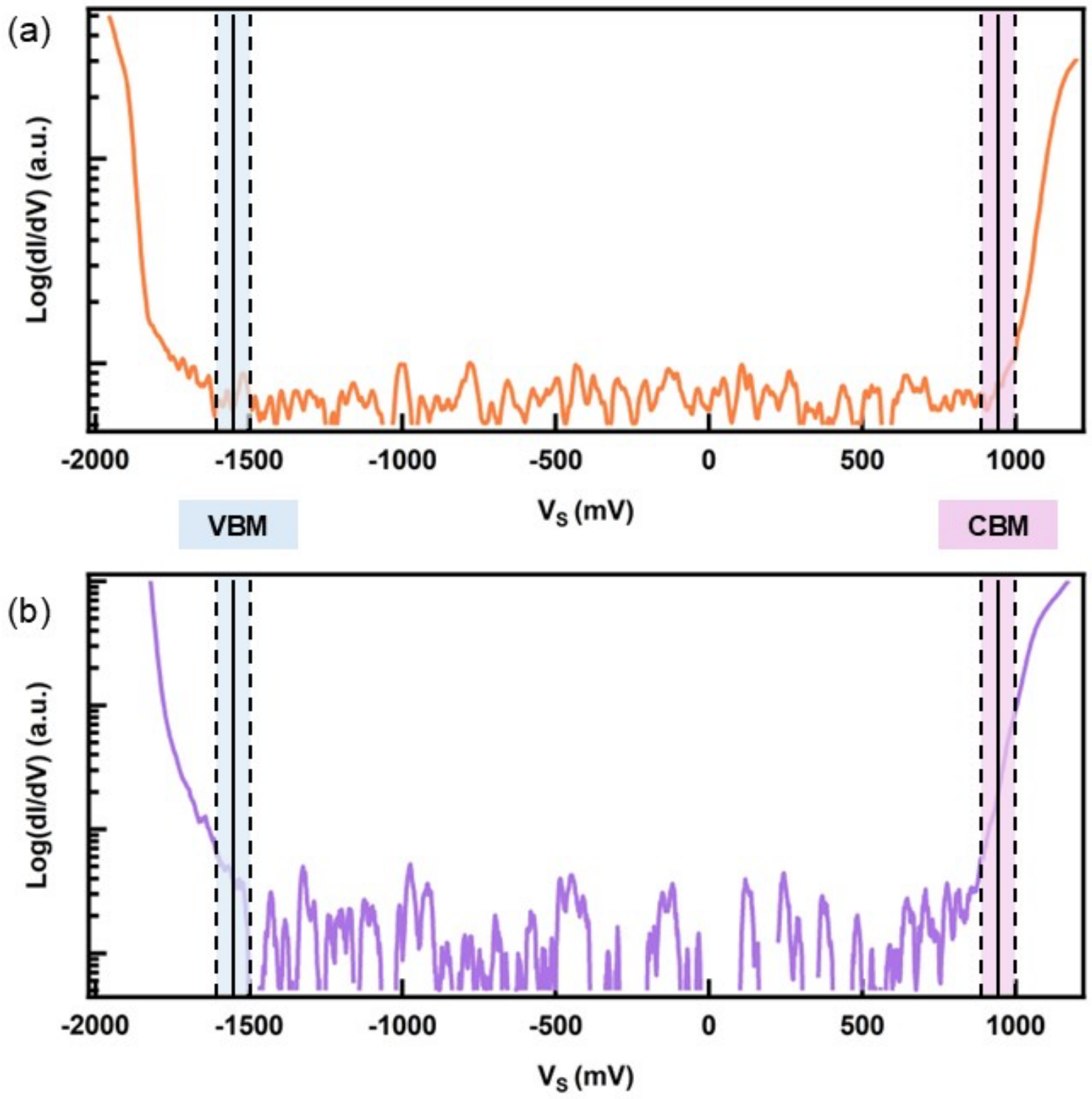

**Figure S1 Determination of Band edges of $WS_2$ from dI/dV spectra.** Data taken on **(a)** pristine $WS_2$ and **(b)** V-$WS_2$. The valence band maximum (VBM, blue shade) occurs at around $V_S = (-1550 \pm 50)$ mV and the conduction band minimum (CBM, pink shade) is at around $V_S = (950 \pm 50)$ mV. The band gap is about $(2.5 \pm 0.1)$ eV.

## S6. Koopmans' condition and intrinsic gap from density functional theory (DFT)

The ionization potential $IP_q$ and electron affinity $EA_{q+1}$ obtained from Kohn–Sham eigenvalues vary linearly with the fraction of exact exchange α. Here we use V substation for tungsten ($V_W$) defect as a probe to enforce the generalized Koopmans' condition between its neutral (q = 0) and singly negatively charged (q = −1) states. For the charged state, the Kohn–Sham eigenvalues of charged supercells were corrected for periodic Coulomb interactions according to

$$\varepsilon_{d,corr}^{KS} = -\frac{2}{q} E_{CCC}, \text{ Eq. (1) [13].}$$

The generalized Koopmans condition $EA_{q+1}=IP_q$ yields α=0.20 (Fig. S2a). With the resulting PBE0 functional, the intrinsic band gap of pristine monolayer $WS_2$ is 2.70 eV without spin-orbit coupling (SOC). Including SOC leads to a direct band gap of 2.40 eV at the K point with strong splitting at VBM (Fig. S2b), which agrees with the experimental band gap of about (2.5 ± 0.1) eV reported in this work.

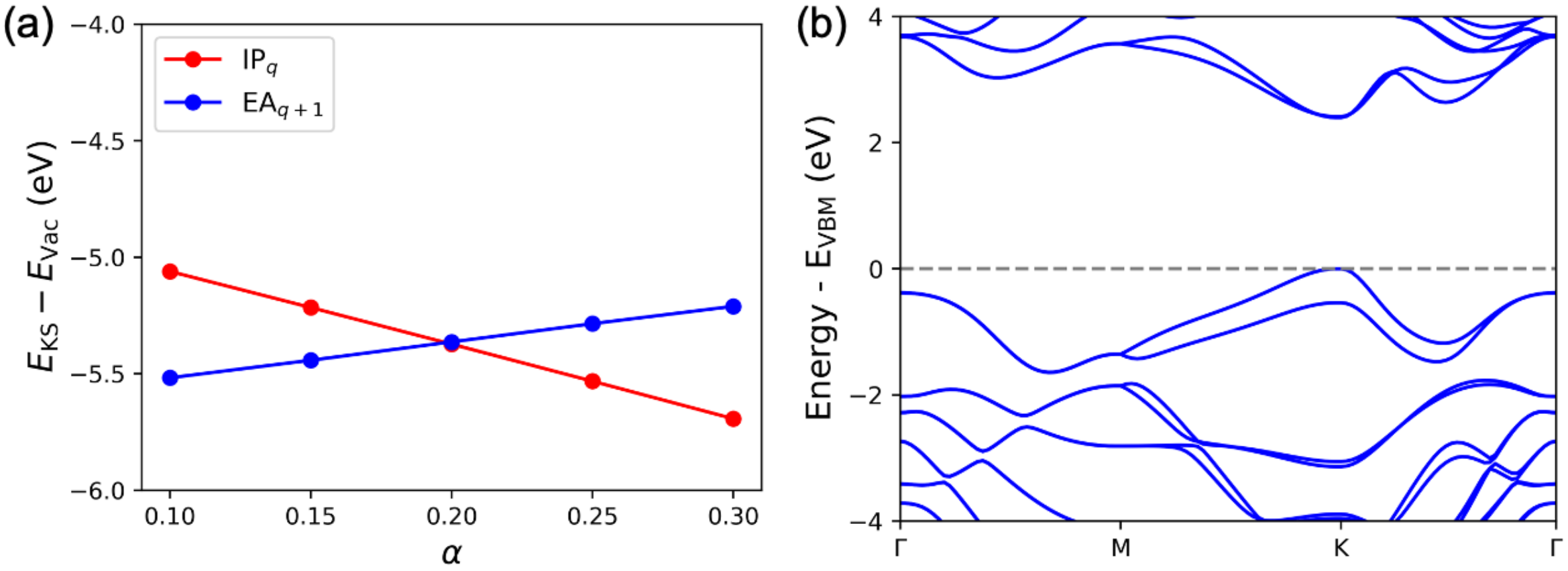

**Figure S2 Band structure of pristine monolayer $WS_2$. (a)** The IP at q=-1 and the EA at q=0 for the defect $V_W$ in ML $WS_2$ as the fraction of exact exchange α. **(b)** Band structure of pristine monolayer $WS_2$ in the primitive cell obtained with the PBE0(α) functional (α = 0.20).

## S7. Thermodynamic charge transition levels

A thermodynamic charge transition level (CTL) is the value of the electron chemical potential at which the stable charge state of a defect changes from q to q+1. CTLs are determined using the total-energy method by equating the formation energies of the two charge states, as shown in the following Eq. (2) [14].

$$\varepsilon_{q+1|q} = E_q^f(\boldsymbol{R}_q) - E_{q+1}^f(\boldsymbol{R}_{q+1}) = E_q(\boldsymbol{R}_q) - E_{q+1}(\boldsymbol{R}_{q+1}) - \varepsilon_F = E_q^{rlx} + \varepsilon_{q+1|q}^{QP}(\boldsymbol{R}_q) \qquad \text{Eq. (2)}$$

where $E_q^f(\boldsymbol{R}_q)$ denotes the defect formation energy at equilibrium structure ($\mathbf{R}_q$) of charge state q. For 2D materials, the energy levels should be aligned to the vacuum level $E_{vac}$, so the Fermi level $\varepsilon_F$ in Eq. (2) is replaced by $E_{vac}$. $E_q^{rlx}$ refers to relaxation energy of charged state while $\varepsilon_{q+1|q}^{QP}$ is the vertical transition level. CTLs obtained from total energies with vacuum-level alignment are relatively insensitive to the choice of exchange–correlation functional; further details are given in Ref. [7].

Figure S3a shows the relative energies of different charge states of the $O_S$ defect with respect to the neutral state (q=0). There are no CTLs within the in-gap range, consistent with no gap states found in experiments and DOS calculations.

Figure S3b shows the relative energies of different charge states of the $V_W$ defect with respect to the neutral state (q=0). We can obtain the defect ionization energies based on the total-energy method discussed earlier for thermodynamic CTLs and band-edge positions relative to $E_{vac}$. The (0/−1) thermodynamic CTL is found at about 0.77 eV above the VBM.

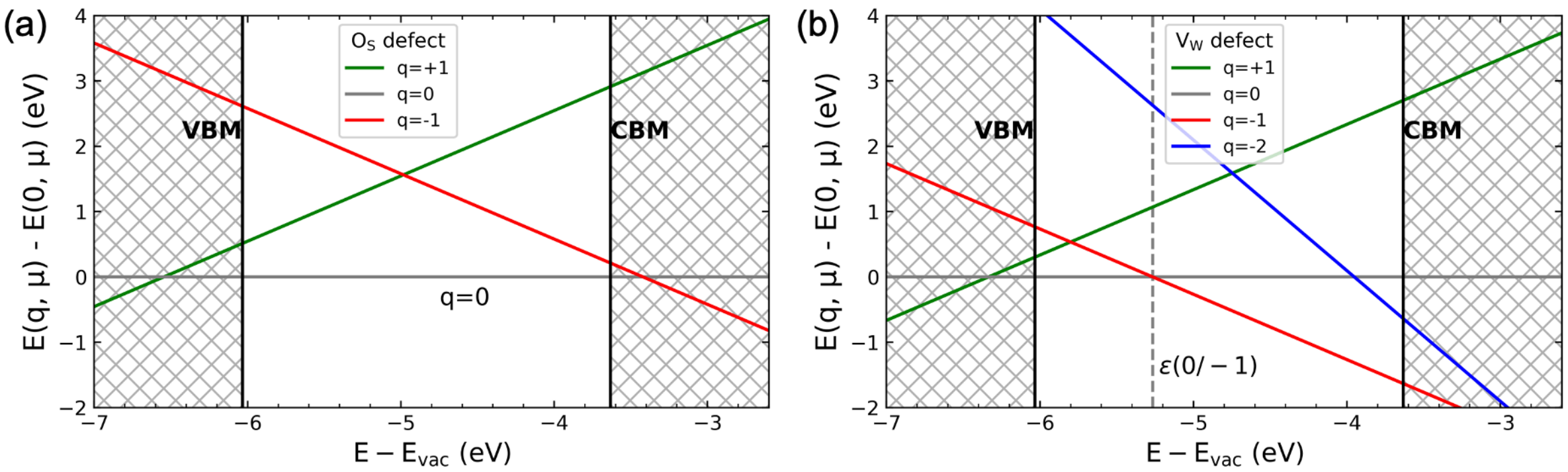

**Figure S3 Relative formation energies of different charged states**. **(a)** $O_S$ defect and **(b)** $V_W$ defect as a function of the Fermi level, where the Fermi level is referenced to the vacuum level.

## S8. Optical spectroscopy on V-$WS_2$ with varying dopant precursor

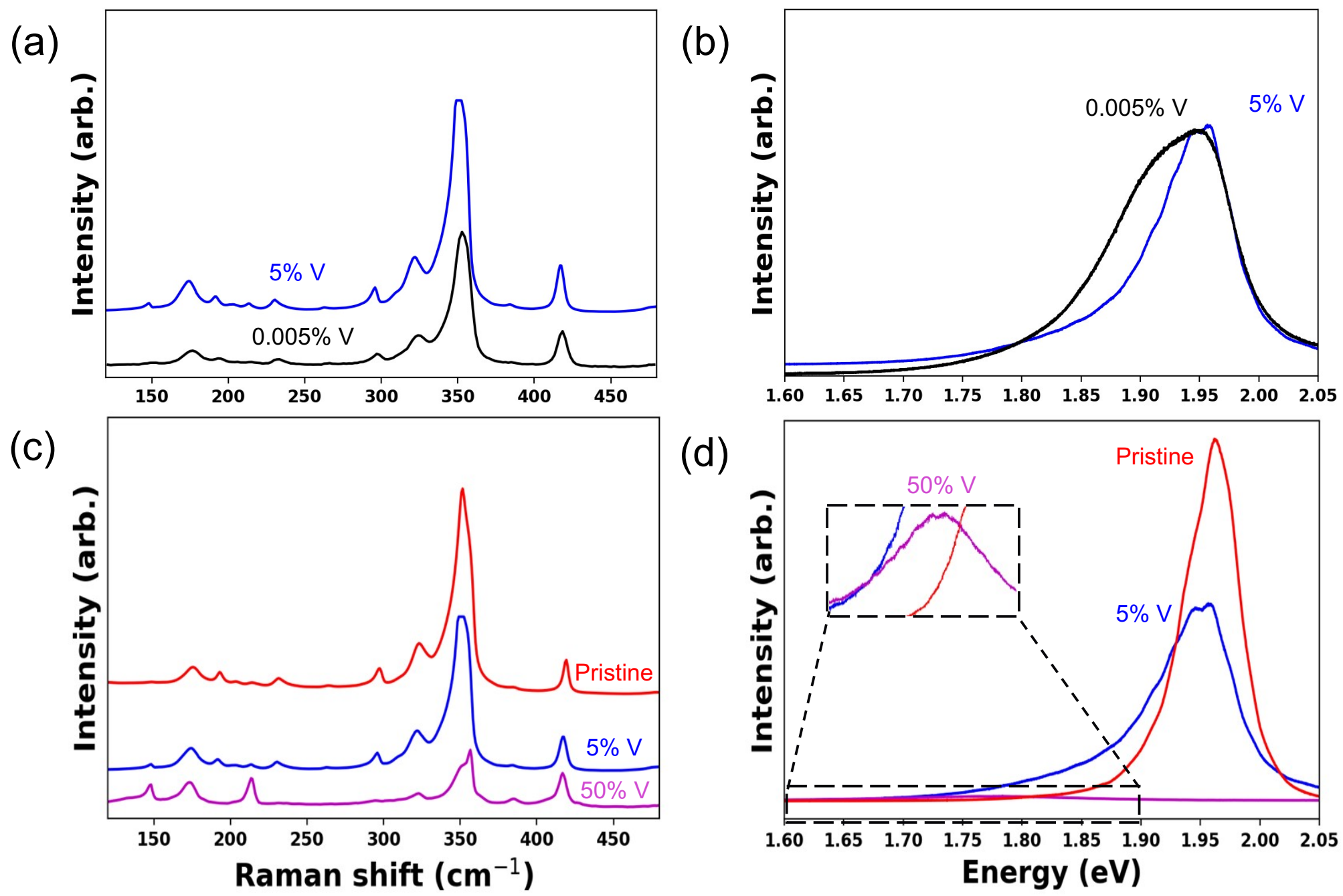

**Figure S4 Optical spectroscopy on V-$WS_2$ with varying dopant precursor. (a)** Raman and **(b)** PL spectra of representative samples with 5% and 0.005% V precursor concentration, respectively. **(c)** Raman and **(d)** PL spectra of pristine $WS_2$ with 5% and 50% V precursor concentration.

## S9. Conductive atomic force microscopy (cAFM) images at negative bias

The cAFM images at negative bias ($-V_S$) collect negative current since electrons are leaving the tip toward the graphite, compared to positive bias where tunneling direction starts at graphite and goes toward the tip. Due to the negative current at negative bias voltage, the data intrinsically presents a larger negative current as darker and a smaller negative current as positive. This means that defects that reduce the current signal will be shown as bright spots and defects that increase the current will be shown as dark spots, as shown in Fig. S5a. To compare the behavior of defects at negative bias, the color scale is inverted to consistently show a reduction in current as dark and an increase in current as bright, as shown in Fig. S5b. This inverted color scale is applied to Fig. S6c and Fig. S6d.

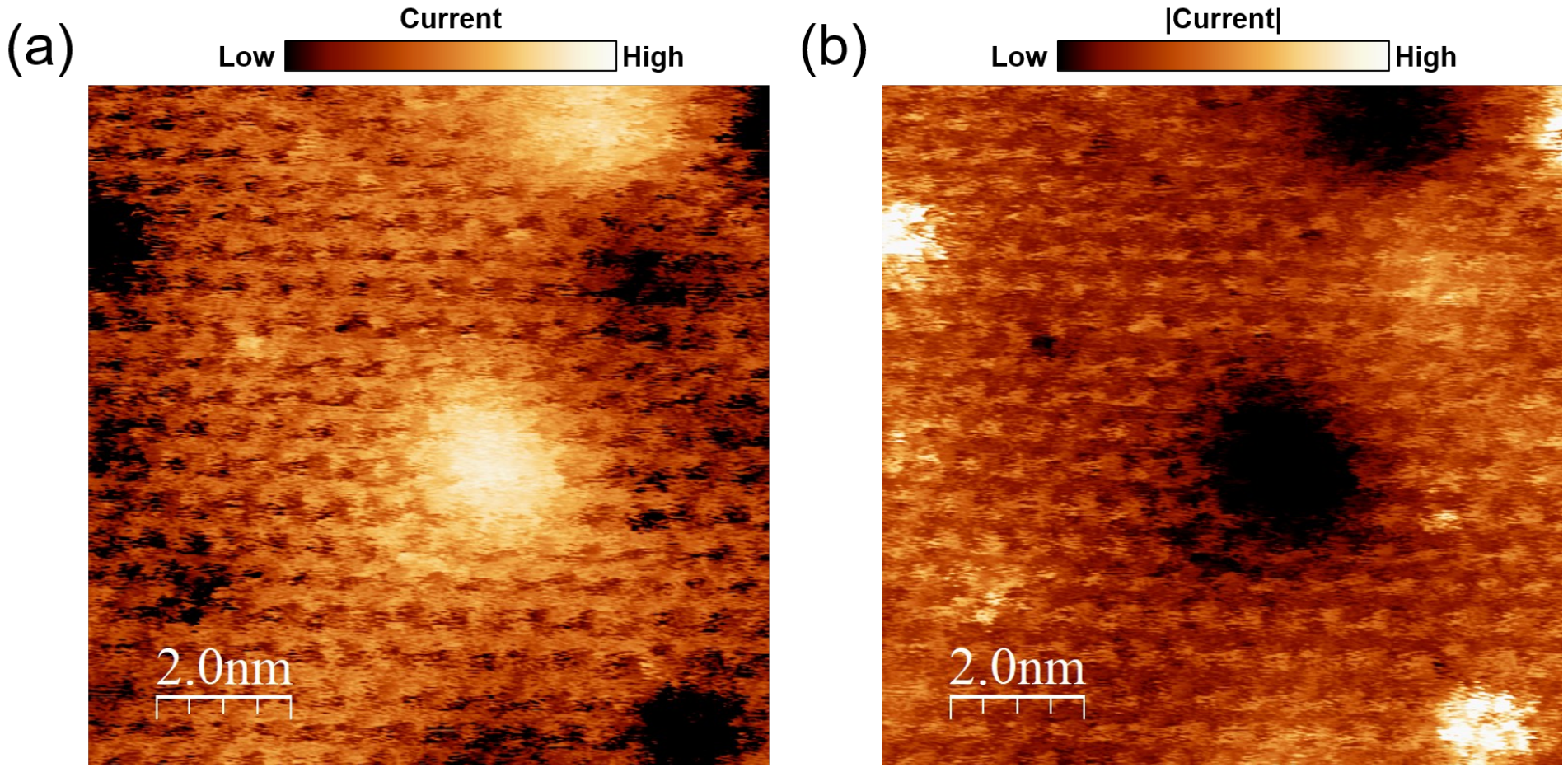

**Figure S5 Absolute versus total current of the same cAFM scan of V-$WS_2$ at negative bias.** (a) A cAFM image presented in absolute value of current, $|I|$. (b) The same cAFM image presented in total current, where negative current is shown as dark. $V_S$= −0.1 V.

## S10. cAFM defect behavior at positive and negative bias

In Fig. S6a and S6b, the defects are imaged at positive bias. This is compared to the same regions imaged at negative bias, using the color scale of absolute value of current explained in section S5, shown in Fig. S6c and S6d, respectively. The cAFM images in Fig. S6 show that the behavior of defects remains consistent at both positive and negative bias.

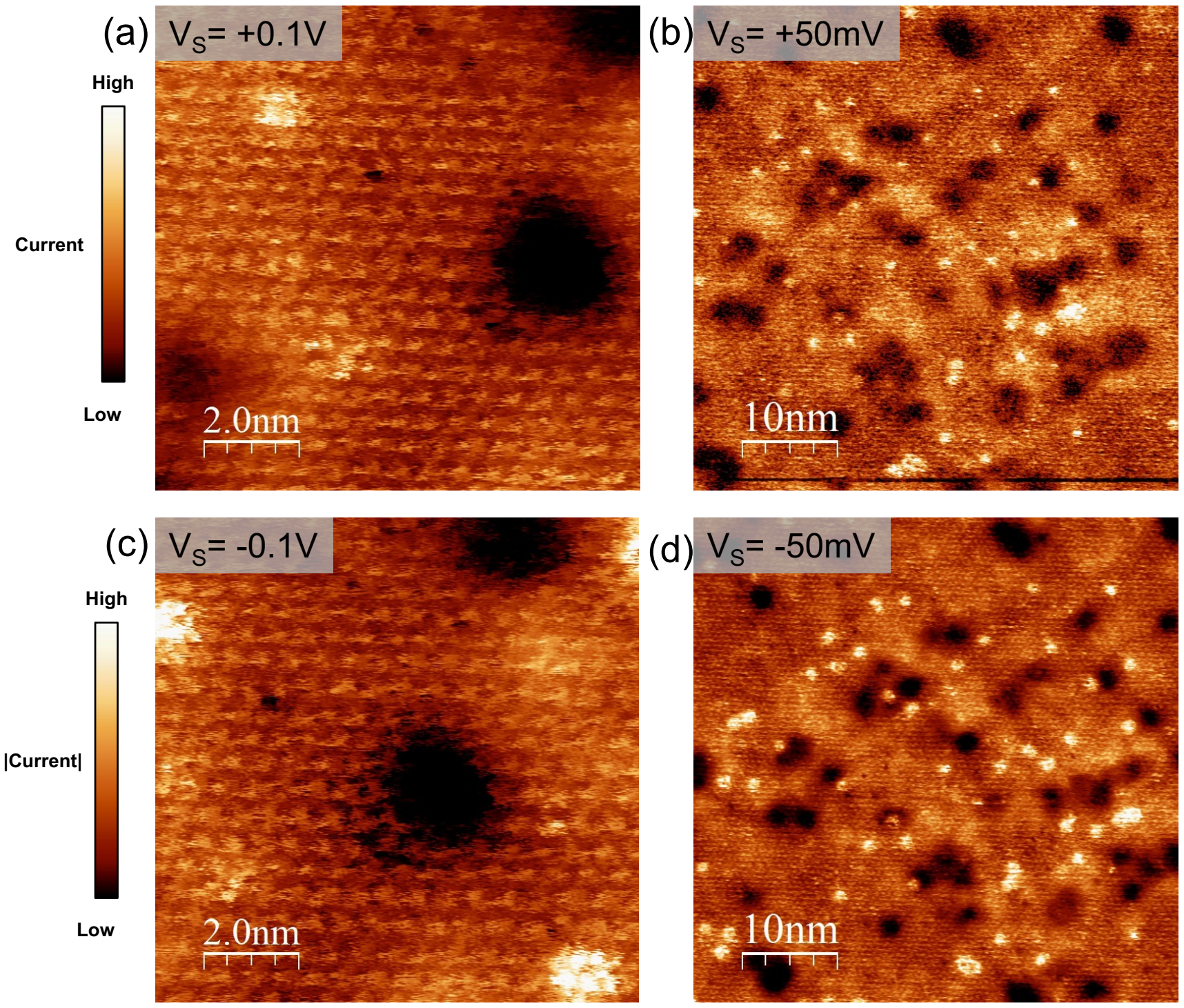

**Figure S6 Comparing cAFM images of V-$WS_2$ defects at $+V_S$ and $-V_S$** (a,b) positive and (c,d) negative bias images of the same region. A comparison between (a) with (c) and (b) with (d) shows that the behavior of the defects is consistent regardless of bias voltage polarity.

## S11. Identification of $CH_S$ defects

$V_W$ and $CH_S$ defects share similarities as they are both negatively charged. Upon close inspection, they can be distinguished based on several unique features in scanning tunneling microscopy (STM) and STS. As shown in Fig. S7a and reported in literature [15,16], $CH_S$ defect also has in-gap states close to VBM but the close-up on the in-gap states displays two smaller peaks followed by one large peak, which is different from the $V_W$ defect. The upward shift in CBM indicates that $CH_S$ is negatively charged. The $dI/dV_S$ image of $CH_S$ defect at the energy of the deepest in-gap state in Fig. S7b shows a hollowed curved triangle with small dots on each side, which is different from the $V_W$ in-gap state.

Figure S8 shows the STM topography image of the same area shown in Fig. 4a, at $V_S = -1.1$ V, which is one example when the two types of negatively charged defects show different features. The $V_W$ defects are circled in blue and the one $CH_S$ defect is circled in red. At this bias, $V_W$ shows three-fold symmetry while $CH_S$ shows up as a bright dot. This provides further evidence that the density of $CH_S$ is noticeably (~one order of magnitude) lower than that of $V_W$.

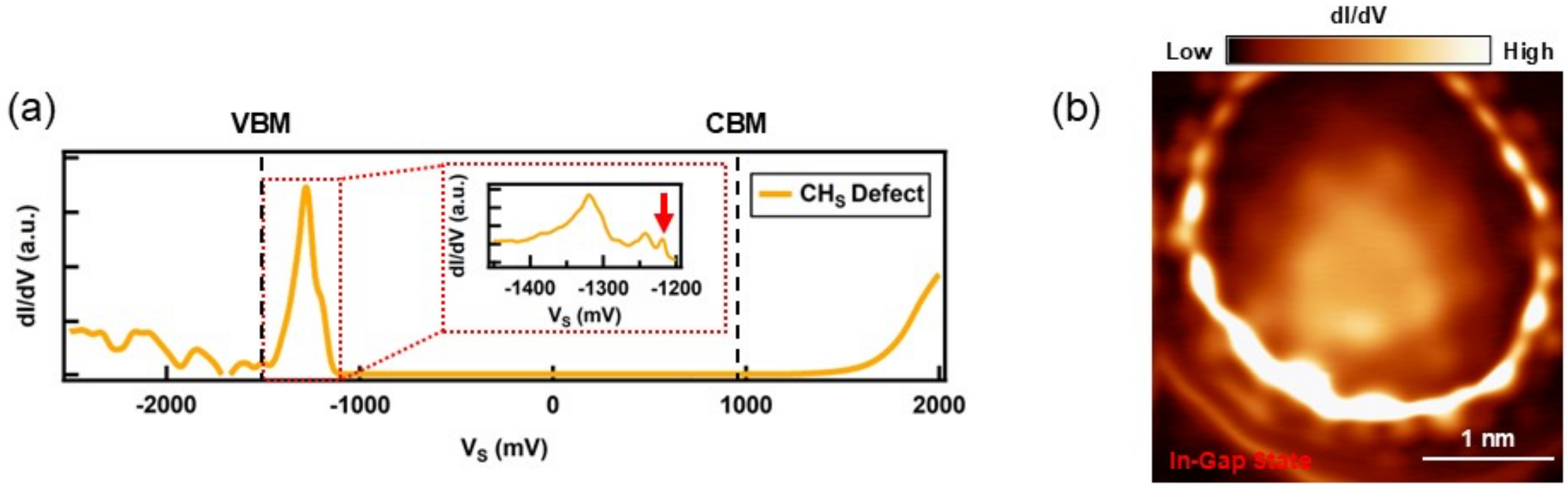

**Figure S7 Spectroscopy signatures of $CH_S$ defect.** (a) dI/dV spectra of an isolated $CH_S$ defect, showing in-gap states close to VBM. Inset zooms in on the in-gap states to show two small peaks followed by one large peak. Shifting of the CBM indicates that the defect is negatively charged. AC excitation is 20mV in (a) and 1mV in the inset. (b) dI/dV map of the deepest in-gap state marked by red arrow in (a), which shows a curved triangle with three small lobes on each side.

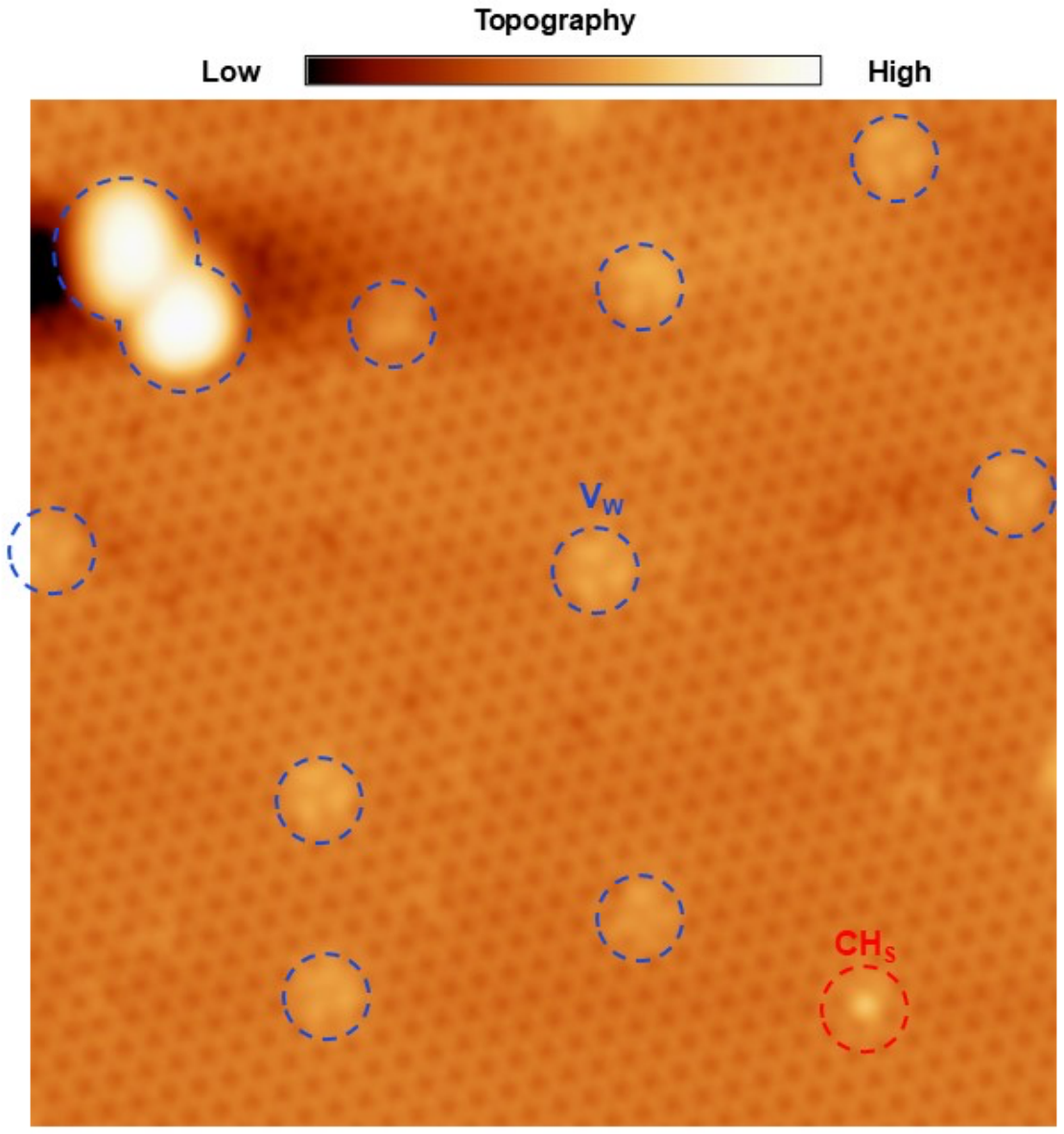

**Figure S8 STM topography map with both $V_W$ and $CH_S$ defects present.** This image is taken in the same region as in Fig. 4(a) in the main text, at $V_S = -1.1$ V. The $V_W$ defects are circled in blue and the one $CH_S$ defect is circled in red. At this bias, $V_W$ shows three-fold symmetry while $CH_S$ shows up as a bright dot.